\documentclass[10pt,twocolumn]{article}
\usepackage[super,sort&compress,comma]{natbib}
\usepackage[paperwidth=210mm,paperheight=297mm,centering,hmargin=2cm,vmargin=2.5cm]{geometry}
\usepackage[font=small,labelfont=bf]{caption}
\usepackage[utf8]{inputenc}
\usepackage[T1]{fontenc}
\usepackage{microtype}
\usepackage{graphicx}
\usepackage{mathbbol}
\usepackage{amssymb}
\usepackage{comment}
\usepackage[usenames,dvipsnames,svgnames,table]{xcolor}
\usepackage[colorlinks,linkcolor=blue,citecolor=blue,urlcolor=blue]{hyperref}
\usepackage{siunitx}
\usepackage{mathtools}
\usepackage{etoolbox}
\usepackage{multibib}
\usepackage{pdfpages}
\usepackage{tikz}
\usepackage{enumitem}
\usepackage{float}

\patchcmd{\thebibliography}{\section*{\refname}}{}{}{} %remove "References" title

\newcites{Met}{ }
\renewcommand{\refname}{}

\newcites{ED}{ }
\renewcommand{\refname}{}

\def\equationautorefname~#1\null{Eq.~(#1)\null}

\protected\def\maxf{\tikz{
    \node[shape=circle, draw=black, fill=white, scale=0.4] {};}}

\protected\def\minf{\tikz{
    \node[shape=circle, fill=black, scale=0.5] {};}}

\protected\def\saddle{\tikz{
    \draw[black, fill=white, semithick, rotate=45] 
        (0, 0) -- (0.06, 0) -- (0.06, 0.06) -- (0, 0.06) -- (0, 0.12) -- 
        (-0.06, 0.12) -- (-0.06, 0.06) -- (-0.12, 0.06) -- (-0.12, 0) -- 
        (-0.06, 0) -- (-0.06, -0.06) -- (0, -0.06) -- cycle;}}

\usepackage{sectsty}
\subsectionfont{\fontsize{9}{15}\selectfont}
\subsubsectionfont{\fontsize{8}{15}\selectfont}

\font\myfont=cmr12 at 13pt
\title{\myfont Topological classification of driven-dissipative nonlinear systems}
    \author{\normalsize Greta Villa$^{1,\ast}$, Javier del Pino$^{1,\ast}$, Vincent Dumont$^{2,3}$, Gianluca Rastelli$^{4}$,
    Mateusz Michałek$^{5}$,\\\normalsize
    Alexander Eichler$^{2,3}$, and  Oded Zilberberg$^1$}
\date{%
    \small
    $^1$Department of Physics, University of Konstanz, 78464 Konstanz, Germany\\
    $^2$Laboratory for Solid State Physics, ETH Z\"{u}rich, 8093 Z\"urich, Switzerland\\
    $^3$Quantum Center, ETH Z\"{u}rich, 8093 Z\"{u}rich, Switzerland\\
    $^4$Pitaevskii BEC Center, CNR-INO and Dipartimento di Fisica, Università di Trento, I-38123, Trento, Italy\\
    $^5$University of Konstanz, Dept. of Mathematics and Statistics,  78457 Konstanz, Germany\\
    $^\ast$ G. V. and J. d. P. contributed  equally to this work.}

\begin{document}
	\maketitle
	\DeclareGraphicsExtensions{.pdf,.png,.jpg}

    \small \textbf{In topology, one averages over local geometrical details to reveal robust global features~\cite{hatcher2002algebraic, Atiyah1990geometry}. This approach proves crucial for understanding quantized bulk transport and exotic boundary effects of linear wave propagation in (meta-)materials~\cite{HasanKane2010, Bernevig2013topological, OzawaRMP2019}. Moving beyond linear Hamiltonian systems, the study of topology in physics strives to characterize open (non-Hermitian)~\cite{Ashida2020review, BergholtzRMP2021} and interacting systems~\cite{Fradkin2013field, Rachel2018interacting}.     
    Here, we establish a framework for the topological classification of driven-dissipative nonlinear systems. Specifically, we define a graph index for the Floquet semiclassical equations of motion describing such systems. The graph index builds upon topological vector analysis theory and combines knowledge of the particle-hole nature of fluctuations around each out-of-equilibrium stationary state.     
    To test this approach, we divulge the topological invariants arising in a micro-electromechanical nonlinear resonator subject to forcing and a time-modulated potential. Our framework classifies the complete phase diagram of the system and reveals the topological origin of driven-dissipative phase transitions, as well as that of under- to over-damped responses. Furthermore, we predict topological phase transitions between symmetry-broken phases that pertain to population inversion transitions. This rich manifesting phenomenology reveals the pervasive link between topology and nonlinear dynamics, with broad implications for all fields of science.} \normalsize
    \vspace{4mm}

\begin{figure*}[ht!]
    \centering
    \includegraphics[width = 1\textwidth]{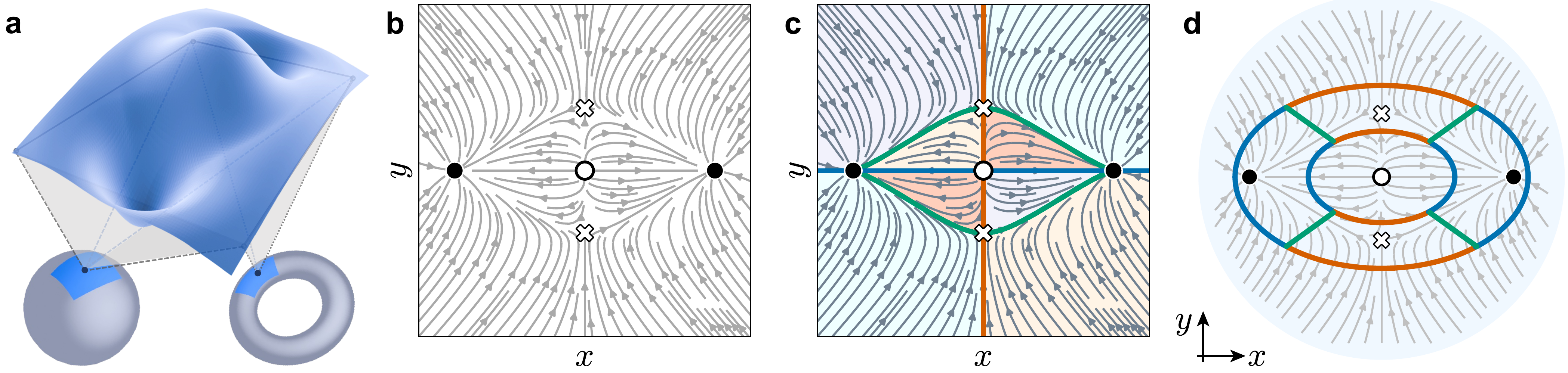}
    \caption{\textbf{Vector flow topology.} \textbf{a.} Topographic landscape corresponding to a 2D surface with a height function. In topology, we average over the local geometric curvature to determine the global shape of a manifold, e.g., a sphere or a torus. \textbf{b.} The topography can be mapped to a gradient vector flow [cf.~Eq.~\eqref{eq:vf}] with peaks, valleys, and saddles, mapping to sources (\protect\maxf), sinks (\protect\minf), and saddles (\protect\saddle) in the flow, respectively. \textbf{c.} A Morse-Smale complex (MSC) splits the manifold into regions (cells) with different behaviour of flow lines. The regions are separated by lines that connect sources to saddles  (red), saddles to sinks (green),  and sources to sinks (blue). \textbf{d.} We can associate the MSC with a colored-graph topological invariant~\cite{Andronov1937,Oshemkov1998}: nodes mark MSC cells, and edges connect between neighbouring cells. The edge colors match the color of the cell boundaries of the MSC. The resulting three-colored graph uniquely encircles regions corresponding to each critical point, and classifies their spatial arrangement. (inset) The graph index is defined on a compact manifold by embedding the flow around the south pole of a sphere~\cite{Andronov1937,Oshemkov1998}.}
    \label{fig:figure1}
\end{figure*}

When we travel on Earth, we experience changes in its surface elevation $f(x,y)$ depending on our local latitude and longitude coordinates $x$ and $y$, respectively. These changes reflect local curvature variations due to mountains, valleys, and passes, see Fig.~\ref{fig:figure1}\textbf{a}. To determine Earth’s global shape, we integrate over the local curvature and obtain a topological index (Euler characteristic) that classifies our planet as a sphere and not a torus~\cite{hatcher2002algebraic, Atiyah1990geometry}. The global shape and its topological invariant are robust to local landscape distortions. Crucially, this ``topology of topography'' goes beyond the classification of global shapes, and can discern between \textit{nonlocal} arrangements of landscape features, e.g., the distribution of watersheds~\cite{Andronov1937,Oshemkov1998,morse2007topology}. The latter discerns which basins are filled with rainfall, as water flows downhill along a gradient flow, 
see Fig.~\ref{fig:figure1}\textbf{b}. By relation of the topography to a 2D vector flow, we can use vector field  topology~\cite{palis2012geometric,gunther2021introduction} to classify the watersheds' arrangement. The classification relies on a Morse-Smale complex (Fig.~\ref{fig:figure1}\textbf{c}) or a related Morse-Smale graph index (Fig.~\ref{fig:figure1}\textbf{d}), that encapsulate the numbers, types, and arrangement of mountains, valleys, ridges and passes. Notably, the flow pattern and the topological index are robust against small perturbations in the topographic structure due to \textit{structural stability}~\cite{Andronov1937}. Vector field topology has wide-ranging applications in various fields, including solar flares, data compression, computer vision, shape analysis, weather forecasting, and robotics~\cite{Longcope_topological_2005, Carlsson2020}.

Topology also plays a key role in wave propagation through media~\cite{hasan2010colloquium}. The transport pertains to bulk energy bands that are characterized by topological indices with associated robust quantized responses and corresponding exotic boundary modes. The topological classification of electron waves has revealed a plethora of novel materials~\cite{bradlyn2017topological}, and prompted their demonstration using a variety of photonic~\cite{OzawaRMP2019}, mechanical~\cite{Shah2024RMP}, electronic~\cite{HasanKane2010}, and atomic~\cite{Cooper2019RMP} metamaterial emulators. As the latter commonly involve driven-dissipative resonators with linear bosonic excitations, recent research deals with the topological classification of open systems with non-Hermitian dynamics related to gain and loss and the associated winding of complex energy bands around exceptional points~\cite{BergholtzRMP2021,Slim2024BKC}. In nonlinear systems, the variety of competing effects and many-body phase transitions makes topological classification much more involved. Commonly, only the linear excitations on top of specific nonlinear many-body states are studied using the approaches detailed above~\cite{Altland2010book,lado2019topological, Xia2020, Mukherjee2020, Mittal2021, Mostaan2022}. Yet, a comprehensive topological classification of the rich combination of driving, dissipation, and many-body interactions remains elusive. 

\begin{figure}[ht!]
    \includegraphics[width = \linewidth]{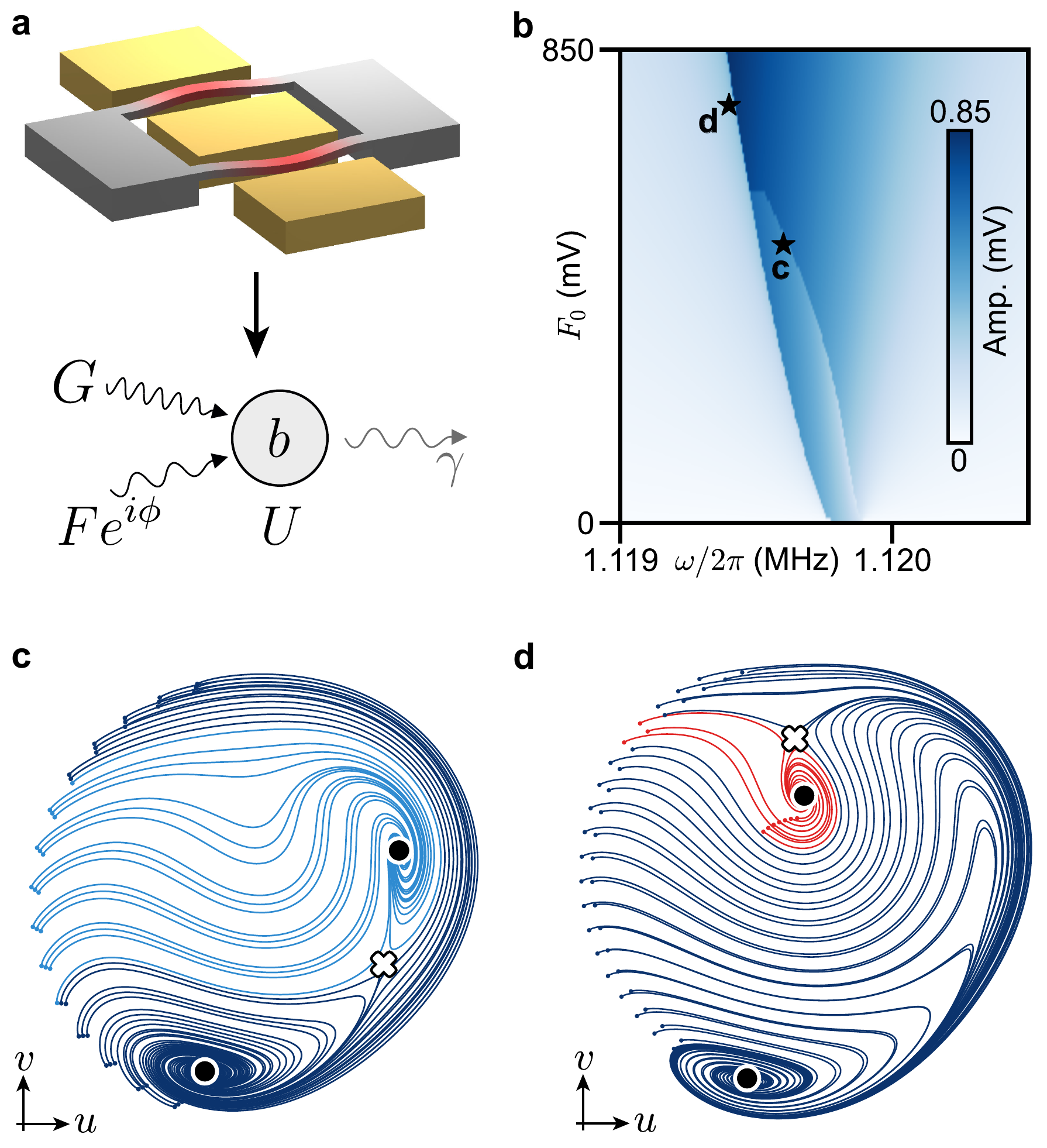}
    \caption{ \textbf{Driven-dissipative NESS of a nonlinear resonator.} \textbf{a.} Top: sketch of the micro-electromechanical resonator used in this work~\cite{Agarwal2008,dumont2024hamiltonian}. The vibrations are driven, and the rotating quadratures $u$ and $v$ at $\omega$ are detected via two electrodes.
    Bottom: model representing Eq.~\eqref{eq:Hamiltonian} depicting a driven-dissipative bosonic mode $b$ subject to one-phonon ($F$) and two-phonon ($G$) drives. The mode has nonlinearity $U$ and experiences phonon loss with rate $\gamma$. \textbf{b.} Stationary amplitude for a red-to-blue frequency sweep of both drives as a function of external resonant drive strength $F_0$, for fixed phase $\phi \approx 0.47\pi$ and parametric drive strength $G_0\approx \SI{3.2}{V}$. \textbf{c.} Rotating frame ringdown reconstruction of the vector flow~\cite{dumont2024hamiltonian} at $F_0 \approx \SI{0.5}{V}$, $\omega/(2\pi) \approx \SI{1.1196}{MHz}$. Blue shades distinguish between clockwise spiralling ringdowns into two different stable NESS. \textbf{d.} Same as \textbf{c} at $F_0 \approx \SI{0.75}{V}$, $\omega/(2\pi) \approx \SI{1.1194}{MHz}$. Blue (red) indicates clockwise (counterclockwise) spirals into different stable NESS; markers are explained in Fig.~\ref{fig:figure3}\textbf{a}. 
    }
    \label{fig:figure2}
\end{figure}

Here, we establish a framework for the topological classification of driven-dissipative nonlinear systems. Specifically, we analyze the Floquet (rotating) semiclassical equations of motion, and use vector field topology to define a graph index that discerns between different arrangements of nonlinear non-equilibrium stationary states (NESS). Our index describes the structural arrangement of attractors (sinks) and repellors (sources) in the rotating phase space, and incorporates the particle-hole characteristics of fluctuations around each NESS. Experimentally, we measure the different graph indexes arising in a micro-electromechanical nonlinear resonator driven by single- and two-phonon drives. Our framework captures the topological origin of locking to different drives, over- to under-damped dissipative transitions, and population inversions in the system. We thus pave the path for a comprehensive unveiling of topological effects in nonlinear media~\cite{shen1984principles,dykman2012fluctuating,Leuch2016,heugel2019classical,parametricBook} and in driven-dissipative collective phenomena~\cite{chitra2015dynamical,Soriente2021,ferri2021emerging,mivehvar2021cavity,hartmann_quantum_2008,Ritsch2013RMP}. 

We first recapitulate the topological analysis of vector fields~\cite{Longcope_topological_2005, Carlsson2020}. Consider, for example, a 2D real vector flow
\begin{equation}\label{eq:vf}
    \dot{\mathbf{r}} = -\nabla f(\mathbf{r}, \boldsymbol{\mu}),
\end{equation}
generated by the gradient of a potential function \(f\), polynomial in coordinates $\mathbf{r}=(x,y)$, with \(\boldsymbol{\mu}\) as additional parameters. Structural stability in the 2D flow~\eqref{eq:vf} requires that~\cite{Andronov1937}:
(i) each trajectory ends in a non-degenerate singular point or a limit cycle, (ii) the flow contains finitely many singular points, i.e., points where $\dot{\mathbf{r}}=0$ (the critical points of $f$), and  (iii) no trajectories exist between saddles. If structural stability holds, trajectories in the flow can be smoothly transformed into each other via perturbations of the parameters $\boldsymbol{\mu}$, while preserving their asymptotic behavior. Parameter changes that preserve the overall flow pattern maintain topological equivalence between the flows. The overall flow pattern is captured by a Morse-Smale complex (MSC)~\cite{wolf2020}, comprising a graph with critical points as nodes, separatrices as edges, and resulting encapsulated cells, see Fig.~\ref{fig:figure1}\textbf{c}.

The MSC can be topologically categorized using a graph index where the edges are three-colored, provided the manifold with coordinates \(\mathbf{r}\) is compact (i.e. closed)~\cite{Oshemkov1998}, see Fig.~\ref{fig:figure1}\textbf{d} and its inset. Isomorphic three-color graphs, meaning that they have matching connectivity and colors, indicate topologically equivalent flows. Such vector flows therefore belong to the same, structurally stable, ``topological phase''. A \textit{topological phase transition} occurs when the graphs become non-isomorphic under a change in $\boldsymbol{\mu}$. This can occur between two different structurally stable flows when (i) the number of NESS (\(N\)) changes due to the formation of unstable fixed points and bifurcations, (ii) the counts of saddles ($N_{\saddle}$), minima ($N_{\minf}$), and maxima ($N_{\maxf}$), change while preserving $N$ (\(N\equiv  N_{\saddle} +  N_{\minf} + N_{\maxf}\)), or (iii) separatrices form or annihilate. Note that parameter changes in \(\boldsymbol{\mu}\) that break the structural stability conditions, e.g., by creating a saddle connection, make the MSC and the graph invariant mathematically ill-defined.

We aim to apply this topological classification framework to driven-dissipative nonlinear systems, relevant for a variety of photonics, mechanics, electric, and cold atoms realizations~\cite{dykman2012fluctuating,parametricBook,hartmann_quantum_2008,Ritsch2013RMP,shen1984principles,Leuch2016,heugel2019classical,chitra2015dynamical,Soriente2021,ferri2021emerging,mivehvar2021cavity}. Here, we present an experiment using a micro-electromechanical resonator (MEMS)~\cite{Agarwal2008,Shaw2016MEMS} under single- and two-phonon drives at frequencies $\omega$ and $2\omega$ and with strengths $F$ and $G$, respectively (Fig.~\ref{fig:figure2}\textbf{a}). Using a lock-in measurement, we retrieve the real quadrature amplitudes $\textbf{q}=(u,v)$ of the resonator relative to a local oscillator at frequency \(\omega\). This establishes a gauge, i.e., a time reference, or alternatively, the axes for $\textbf{q}$ in the rotating phase space. We measure the stationary amplitude of the system for a fixed $F$ while slowly varying $\omega$ and then stepping $G$, see Fig.~\ref{fig:figure2}\textbf{b}. Sharp features correspond to first-order phase transitions between different NESS~\cite{Bartolo2016, Leuch2016, heugel2019quantum,beaulieu2023observation}. A parameter sweep generally samples a subset of available stationary phases but does not capture the complete rotating vector flow of the system. For any particular set of parameters, however, we can reconstruct the time-evolution of \(u\) and \(v\) using several ringdown measurement with variable initial conditions~\cite{dumont2024hamiltonian}, see dots and experimental trajectories in Figs.~\ref{fig:figure2}\textbf{c} and \textbf{d}. These trajectories are analogous to topographic vector flows (Fig.~\ref{fig:figure1}\textbf{c}), but in a rotating-frame phase space.

In the following, we will construct a graph invariant to classify the measured vector flows and obtain a topological phase diagram for the system. We note, however, that the measured vector flows differ from gradient flows introduced in Fig.~\ref{fig:figure1}. Specifically, the flows observed in Fig.~\ref{fig:figure2} (i)~have only saddles as unstable fixed points, (ii)~lack sources, and (iii)~feature sinks with clockwise/counterclockwise spirals, corresponding to faster/slower dynamics than the reference oscillator. To assess the topological significance of the observed differences, we write the effective Hamiltonian of the system in a rotating frame at frequency $\omega$, using a recent Floquet expansion technique based on counting driving phonons~\cite{kovsata2022fixing,seibold2024floquet} ($\hbar=1$),
\begin{align}
H  =&\left(-\Delta+U\right)b^{\dagger}b+\frac{U}{2}b^{\dagger2}b^{2}+\frac{G}{2}\left(b^{2}+b^{\dagger}\phantom{}^{2}\right)\nonumber\\
&-F\left(b^{\dagger}e^{-i\phi}+b e^{i\phi}\right)\,.
\label{eq:Hamiltonian}
\end{align}
Here, the operator $b$ annihilates phonons in the resonator, such that \(\langle b \rangle = \sqrt{m \omega / 2 } (u + i v / (m \omega))\), $U$ parametrizes  phonon-phonon interactions, and $\Delta = (\omega^2 - \omega_0^2)/(2  \omega)$ is the detuning between the natural frequency of the resonator $\omega_0/(2\pi)\approx \SI{1.1198}{MHz}$ and drive frequency $\omega$, cf. schematic in Fig.~\ref{fig:figure2}\textbf{a}. The phase $\phi$ is the difference between the phases of the single- and two-phonon drives. 
The experiment operates in the semiclassical limit due to weak interactions (\(U\)) and high phonon population (\(\langle b^{\dagger}b\rangle \gg 1\)), allowing us to use the mean-field approximation \(\langle AB \rangle \approx \langle A \rangle \langle B \rangle\). Additionally, we add the phonon dissipation rate \(\gamma\), such that the rotating quadratures follow the flow, 
\begin{align}~\label{eq:our_flow}
\dot{\mathbf{q}} = -J\nabla \tilde{H}(\mathbf{q})- \nabla R(\mathbf{q}),&& \text{where }\, J = \begin{bmatrix}
0 & -1 \\
1 & 0  
\end{bmatrix},
\end{align}
and the semiclassical Hamiltonian potential $\tilde{H}(\mathbf{q})$ is obtained by replacing $b\rightarrow\langle b\rangle$, in Eq.~\eqref{eq:Hamiltonian}. In the rotating frame, dissipation enters as a Rayleigh potential of the form $R(\mathbf{q})= \gamma |\mathbf{q}|^2/4$.

\begin{figure*}[ht!]
    \centering
    \includegraphics[width = \textwidth]{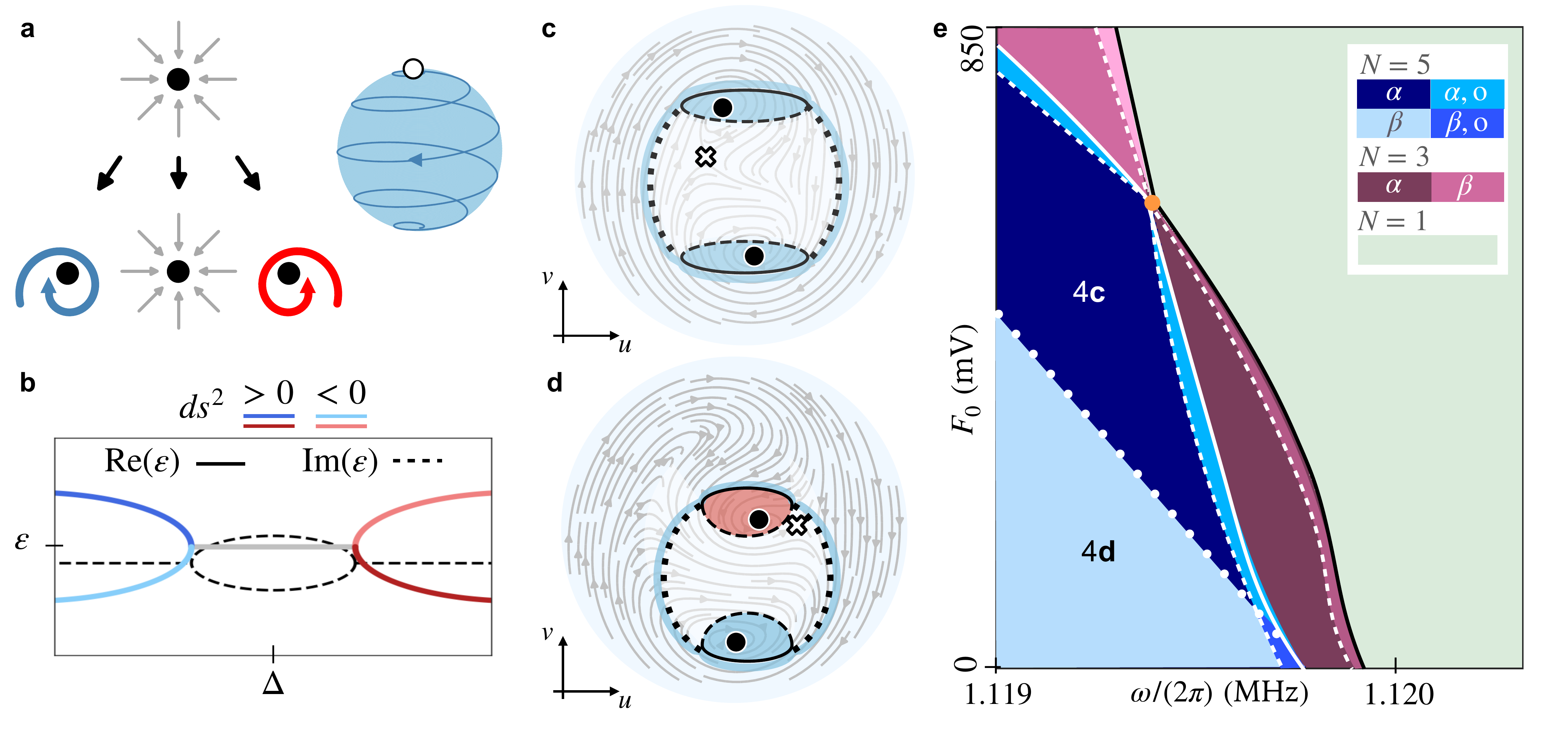}
    \caption{\textbf{Topological invariant for driven-dissipative nonlinear systems.} 
    \textbf{a.} Sinks are mapped to attractive foci with positive (blue), negative (red), or no chirality (no color). We embed the flow on a sphere using an inverse stereographic projection \((u,v) \mapsto \left(2ur^2, 2vr^2, r(X^2 - r^2)\right)/(X^2 + r^2)\), , with $X=\sqrt{u^2+v^2}$, on a three-coordinate axis. A large sphere radius \(r\) ensures all critical points are in the southern hemisphere. Nonlinearity creates a dominant flow at high amplitudes, acting as a clockwise source (\(U < 0\)) at the north pole. \textbf{b.} Bogoliubov excitation spectrum, $\epsilon$, around a NESS. Particle- (hole-)like excitations, discerned by a positive (negative) symplectic norm $ds^2$, correspond to clockwise (counterclockwise) chiralities~\cite{Soriente2020,dumont2024hamiltonian}. Exceptional points (collapses in the complex eigenexcitation spectrum) mark transitions between underdamped and overdamped excitations. \textbf{c.}\textbf{d.} The graph index for topologically equivalent flows as in Fig.~\ref{fig:figure2}\textbf{c},\textbf{d}, respectively. Edge markers exchange colors in Fig.~\ref{fig:figure1}\textbf{d}: dotted(saddle to ``virtual'' source), dashed  (saddle to sink), solid (sink to virtual source), replacing the old red, blue, and green, respectively.
    Cell colors mark local chirality, and the halo around the graphs highlights background chirality.\textbf{e.} Full phase diagram of the system in Eq.~\eqref{eq:Hamiltonian} via our topological classification framework. Hue indicates solution count; shades denote chirality or separatrix connectivity (see legend). Solid, dashed, and dash-dotted lines represent transitions in solution number, exceptional points, and separatrix connectivity changes, respectively. 
    For clarity, the width of the overdamped regions are exaggerated.}
    \label{fig:figure3}
\end{figure*}

In the symplectic limit (\(\gamma = 0\)), the flow~\eqref{eq:our_flow} exhibits closed orbits orthogonal to \(\tilde{H}\)'s gradient flow. In the dissipative  Rayleigh (\(\tilde{H} = 0\)) limit, the flow has a single sink at the origin. While standard Morse-Smale topological classification applies to the gradients flows of \(\tilde{H}\) or of \(R\) independently, the combination in Eq.~\eqref{eq:our_flow} cannot be mapped to a gradient-type flow. Consequently, sinks and sources in the flow are not straightforwardly linked to the critical points of the generating potentials. Indeed, sinks in Figs.~\ref{fig:figure2}\textbf{c},\textbf{d}  form \textit{both} at the maxima and minima of \(\tilde{H}\) and can exhibit different flow chiralities, see Fig.~\ref{fig:figure3}\textbf{a}.

The local chirality of the flow around each NESS in Fig.~\ref{fig:figure2} is linked to the particle-hole character of excitations~\cite{Soriente2020,Soriente2021,dumont2024hamiltonian}, see Fig.~\ref{fig:figure3}\textbf{b}. Specifically, eigenstates of the Bogoliubov-de Gennes Hamiltonian around each NESS are ascribed a vector norm weighed with the metric $J$ in Eq.~\ref{eq:our_flow}. This so-called symplectic norm, $ds^2$, takes values \(\mathrm{sign}(ds^2)\in\{1,\varnothing, -1\}\) that encode particle, overdamped, and hole excitations, respectively. Note that particle and hole excitations correspond to clockwise and counterclockwise chiralities~\cite{dumont2024hamiltonian}. Overdamped excitations appear after an exceptional point with the two lifetimes corresponding to squeezed and antisqueezed resonator quadratures, respectively~\cite{huber2020spectral}, see Fig.~\ref{fig:figure3}\textbf{b}.  Yet, the MSC-based invariant in Fig.~\ref{fig:figure1}\textbf{d} overlooks this aspect, as the chirality of spirals can be flipped by a mirror reflection without altering the graph index. As the excitations' nature is fixed by their relation to the reference oscillator (a gauge fixing), the latter restricts the allowed deformations of $\boldsymbol{\mu}$. As such, analogous to symmetry-protected topological phases~\cite{senthil2015symmetry}, we will topologically classify structurally stable vector flows under deformations that respect the chirality around all critical points. Mathematically, this implies that the allowed deformations must commute with transformations that change the symplectic norm  around any  NESS. 

We encode the added symmetry constraint given by the symplectic norm into the graph invariant (cf.~Fig.~\ref{fig:figure1}\textbf{d}) by the following construction: (i)~We distinguish between sinks using colors based on \(\mathrm{sign}(ds^2)\),  cf.~Fig.~\ref{fig:figure3}\textbf{a}; (ii)~To handle flow lines extending to (from) infinity in the $\mathbf{q}$ plane, we embed the flow on a sphere, mapping these lines to a ``virtual source'' at the north pole, see Fig.~\ref{fig:figure3}\textbf{a}.
Here, the dissipation potential \(R\) creates streams from the north to the south pole with a chirality dictated by the nonlinearity $U$, which we also decorate by colors and markers; (iii)~The MSC complex, its line colors, and the ensuing graph index are built using the same rules as in Fig.~\ref{fig:figure1}\textbf{d}, where we recall that we only have sinks and saddles in the presence of a single virtual source. For distinction, instead of three colors, we use three line-dashing styles; (iv) We introduce colored faces to the graph index, indicating each of the NESS's chirality \(\mathrm{sign}(ds^2)\). Two topological phases are equivalent if their graphs are isomorphic, preserving face colors, nodes, and edge dashing. We thus find that the measured flows in Figs.~\ref{fig:figure2}\textbf{c} and \textbf{d} are topologically distinct, see Figs.~\ref{fig:figure3}\textbf{c} and \textbf{d}, respectively. Crucially, this distinction is commonly overlooked when counting only the total number $N$ of stationary states.

\begin{figure*}
    \centering
    \includegraphics[width =\textwidth]{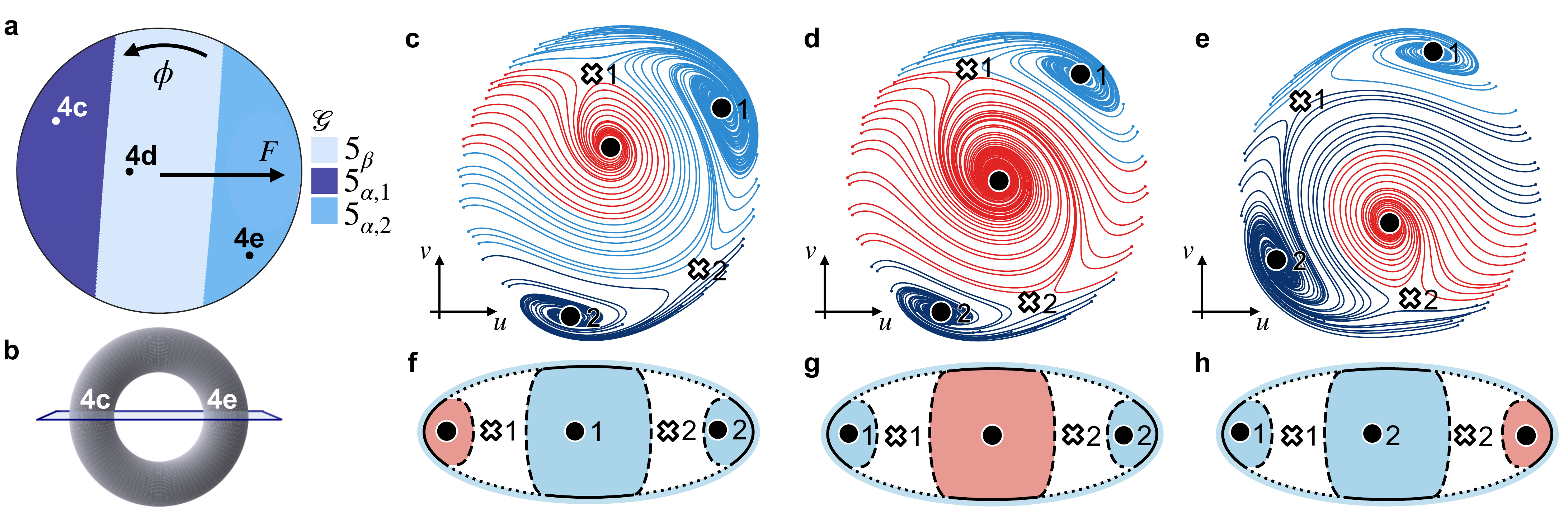}
    \caption{\textbf{Topological phase transition in vector field connectivity.} \textbf{a.} A zoom in on the region with $N=5$ solutions in Fig.~\ref{fig:figure3}\textbf{e} as a function of $F_0 \in (0.00, 4 \times 10^{-4})$ and $\phi \in (0, 2\pi)$. \textbf{b.} The transition between $5_{\alpha1}$ and $5_{\alpha2}$ via  $5_\beta$ resembles a topological \(\mathbb{Z}_2\) transition as the fixed time reference fixes a gauge. This is akin to forcing a transition between two sides of a torus when moving on the cut plane. \textbf{c.}-\textbf{e.} Experimentally measured flows of the phases $5_{\alpha1}$,  $5_\beta$, and $5_{\alpha2}$, respectively. For each panel,  $F_0 = \SI{500}{mV}$, $F_0 = \SI{150}{mV}$ and $F_0 = \SI{-500}{mV}$ respectively, while we keep $\omega/2\pi = \SI{1.1194}{MHz}$ (equivalent to a mirror reflection between \textbf{c} and \textbf{e}). \textbf{f.}-\textbf{h.} Associated graph indexes to the flows \textbf{c}-\textbf{e}. }
    \label{fig:figure4}
\end{figure*}

Equipped with our new graph index, we can now predict the system's complete phase diagram, see Fig.~\ref{fig:figure3}\textbf{e}. It is generally split into three regions according to the number of critical point solutions: $N=1$ (1 stable solution), $N=3$ (2 stable, 1 unstable), and $N=5$ (3 stable, 2 unstable). Different topological phases split the diagram further into 
subparts, which we enumerate using greek-letter subscripts, e.g., $N_\alpha$. Topological phase transitions occur at all marked lines, corresponding to three types of possible effects: (i)~solid lines indicate transitions where the variety of NESS (\( N_{\saddle},  N_{\minf}, N_{\maxf}\)) changes; (ii) dashed lines mark under- to over-damped transitions around a NESS; and (iii) dotted lines mark graph connectivity changes that do not involve any of the changes (i) and (ii). Already by the number of solutions, we identify that regions $N=1$ and $N=5$ correspond to a standard driven harmonic oscillator phase, and to a coexistence region between low- and high-amplitude phases in response to both the drives \(F\) and \(G\)~\cite{Leuch2016,heugel2019quantum,parametricBook,boness2024resonant}. For $N=3$, our classification draws a distinction between $3_\alpha$ and $3_\beta$ regions, signalling the graph-index variety exposed in Figs.~\ref{fig:figure3}\textbf{c} and \textbf{d}. In other words, our flow topology distinguishes between the Kerr Parametric Oscillator (KPO, $3_\alpha$) and a Kerr Driven Oscillator (KDO, $3_\beta$). Physically, in $3_\alpha$, the resonator locks onto the parametric drive $G$, with flows showing two sinks of equal chirality split by a saddle. The flow is only  perturbatively $\mathbb{Z}_2$-symmetry broken~\cite{Papariello2016,Leuch2016} by $F$. In $3_\beta$, a standard Duffing bistable regime appears in response to $F$, with $G$ acting as a perturbation.
Here, the resonator locks onto the single-phonon drive, with flows showing two sinks of opposite chirality and a saddle. To maintain the topological distinction, all four phase boundaries converge at a multi-critical point at the crossing between the solid lines.

Our graph invariant captures changes in the non-Hermitian excitations around \textit{each} NESS, cf. flipped chirality between $3_\alpha$ and $3_\beta$. The dashed boundaries in Fig.~\ref{fig:figure3}\textbf{e},  capture further transitions driven by the interplay between damping and squeezing (gain); the latter arising  from nonlinearity and two-phonon driving. Specifically, the interplay can modify the excitation around a NESS from being underdamped (\(\tilde{H}\) flow dominates) to overdamped (\(R\) flow dominates). This  causes spirals to collapse into attracting nodes, and lose chirality. Thus, cells in the graph index shift from colorful (finite chirality) to colorless (no chirality). In other words, the dashed lines mark the exceptional points of the complex fluctuation eigenmodes around certain NESS, spanning a finite region between these points and the NESS instability at Fig.~\ref{fig:figure3}\textbf{b}. As such, the dashed lines appear around the solid transition lines that change $N$ (Fig.~\ref{fig:figure3}\textbf{e}).  Note that this distinction allows us to describe dissipation-induced phase transition, where saddles become attracting nodes due to strong damping~\cite{Flynn2020,Soriente2021,ferri2021emerging,parametricBook}. We denote distinct overdamped regions originating from different spiraling graphs with an additional subscript $N_{{\alpha},{o}}$.

Last, we switch our focus to the $N=5$ region in the phase diagram in Fig.~\ref{fig:figure3}\textbf{e}, where two phases $5_\alpha$ and $5_\beta$ are separated by a dotted line. Here, the vector flow has two clockwise and one counterclockwise sinks, as well as two saddles. Interestingly, the dotted line marks a topological phase transition that occurs only through changes in graph connectivity. To better understand the effect, we zoom in on the transition region and vary $F$ and $\phi$, see Fig.\ref{fig:figure4}\textbf{a}. We observe that $5_\alpha$ transitions into $5_\beta$, only to return back into a phase that resembles $5_\alpha$, but with a $\pi$-shifted phase space. Analogous to crossing a torus along a fixed cut (Fig.~\ref{fig:figure4}\textbf{b}), this highlights a \(\mathbb{Z}_2\) distinction between the two $\pi$-shifted regions, which we denote $5_{\alpha1}$ and $5_{\alpha2}$. To visualize the different phases, we experimentally reconstruct flows from each of the phases, see Figs.~\ref{fig:figure4}\textbf{c}-\textbf{e}. The measured flows demonstrate that graph-index transitions are caused by a rearrangement in the separatrices in the vector flow. At large positive (negative) \(F\) (Figs.~\ref{fig:figure4}\textbf{c},\textbf{e}), the flow exhibits that the counterclockwise sink is connected only to saddle \(\saddle_1\) (\(\saddle_2\)).  At \(F \approx 0\) instead (Fig.~\ref{fig:figure4}\textbf{d}), both saddles are connected to the counterclockwise sink. We draw the flows' corresponding graphs in Figs.~\ref{fig:figure4}\textbf{f}-\textbf{h}. Indeed, the graphs defining $5_{\alpha1}$ and $5_{\alpha2}$  are related by a \(\mathbb{Z}_2\) symmetry \((u,v) \mapsto -(u,v)\), which inverts the NESSs through the origin of phase space. Note that the fixed gauge in the experiment makes saddles \(\saddle_{1,2}\) distinguishable, rendering graphs in Figs.~\ref{fig:figure4}\textbf{f} and \textbf{h}  non-isomorphic. The \(\mathbb{Z}_2\) transition between these graphs appears as an inversion of the dominant basins of attraction for phonon population in phase space. The latter can serve as an experimental fingerprint for such topological phase transitions. In other words, such a $\mathbb{Z}_2$ transition will lead to observable changes in noise activation rates between attractors in thermal or quantum regimes~\cite{dykman2012fluctuating}.

Our work elucidates how topology manifests in nonlinear and non-Hermitian driven systems, establishing a theory beyond existing methodologies tailored for linear systems~\cite{OzawaRMP2019, delPino2022NH}. We harness vector field topology to classify the equations of motion arising from the competition between conservative and dissipative forces. Experimentally, we demonstrate topological flows in a driven micro-electromechanical resonator, and reveal an unexpected population-inversion topological transition. Our unique approach paves the way for the exploration of topological effects in quantum circuits~\cite{seibold2024floquet}, many-body collective phenomena in cold atoms~\cite{Soriente2021,ferri2021emerging}, nonlinear phonon networks~\cite{Dai2024,Heugel2024}, and nonlinear optics~\cite{shen1984principles}. The classification of the many-body flows in such high-dimensional phase-spaces will rely on extensions of Morse-Smale theory~\cite{Carlsson2020}. An interesting open question pertains to how the topology is impacted by stochastic classical or quantum noise that introduces uncertainty to the flow lines. Furthermore, the flow topology can significantly impact our understanding and control of stochastic noise-induced activation pathways in a variety of fields, including neuromorphic computing~\cite{Markoviv2024},  biological information processing~\cite{mcdonnell2009stochastic}, climate physics~\cite{ragone2018computation} and machinery fault detection~\cite{qiao2019applications}. Similarly, we expect applications in nonlinear sensors~\cite{Gammaitoni2002,rodriguez2020enhancing}, active matter systems~\cite{Marchetti2013RMP} and bosonic error correction codes~\cite{terhal2020towards}.

We acknowledge support from ETH Zurich through the Postdoctoral Fellowship Grant No. 23-1 FEL-023, Swiss National Science Foundation (SNSF) through the Sinergia Grant No.~CRSII5\_206008/1, and the  Deutsche Forschungsgemeinschaft (DFG) via project numbers 449653034 and number 467575307, and through SFB1432. We thank  Nicholas E. Bousse and T. W. Kenny for providing the MEMS device.

Author contributions: G.~V., J.~d.~P., M.~M., and O.~Z. developed the theoretical framework. V.~D. and A.~E. designed the experiment, V.~D. performed the measurements. G.~R. aided in analyzing the resonator model. A.~E. and O.~Z. conceived the project and coordinated the work. All authors contributed to the interpretation of the results and to writing the manuscript.

% Reset counters for multibib references
\setcounter{enumiv}{0}
% Main bibliography
\bibliographystyle{naturemag}
{\small
\bibliography{references}}

\begin{thebibliography}{10}
\expandafter\ifx\csname url\endcsname\relax
  \def\url#1{\texttt{#1}}\fi
\expandafter\ifx\csname urlprefix\endcsname\relax\def\urlprefix{URL }\fi
\providecommand{\bibinfo}[2]{#2}
\providecommand{\eprint}[2][]{\url{#2}}

\bibitem{hatcher2002algebraic}
\bibinfo{author}{Hatcher, A.}
\newblock \emph{\bibinfo{title}{Algebraic topology}}
  (\bibinfo{publisher}{Cambridge University Press}, \bibinfo{year}{2002}).

\bibitem{Atiyah1990geometry}
\bibinfo{author}{Atiyah, M.~F.}
\newblock \emph{\bibinfo{title}{The geometry and physics of knots}}
  (\bibinfo{publisher}{Cambridge University Press}, \bibinfo{year}{1990}).

\bibitem{HasanKane2010}
\bibinfo{author}{Hasan, M.~Z.} \& \bibinfo{author}{Kane, C.~L.}
\newblock \bibinfo{title}{Colloquium: Topological insulators}.
\newblock \emph{\bibinfo{journal}{Rev. Mod. Phys.}}
  \textbf{\bibinfo{volume}{82}}, \bibinfo{pages}{3045--3067}
  (\bibinfo{year}{2010}).

\bibitem{Bernevig2013topological}
\bibinfo{author}{Bernevig, B.~A.}
\newblock \emph{\bibinfo{title}{Topological insulators and topological
  superconductors}} (\bibinfo{publisher}{Princeton university press},
  \bibinfo{year}{2013}).

\bibitem{OzawaRMP2019}
\bibinfo{author}{Ozawa, T.} \emph{et~al.}
\newblock \bibinfo{title}{Topological photonics}.
\newblock \emph{\bibinfo{journal}{Rev. Mod. Phys.}}
  \textbf{\bibinfo{volume}{91}}, \bibinfo{pages}{015006}
  (\bibinfo{year}{2019}).

\bibitem{Ashida2020review}
\bibinfo{author}{Yuto~Ashida, Z.~G.} \& \bibinfo{author}{Ueda, M.}
\newblock \bibinfo{title}{Non-hermitian physics}.
\newblock \emph{\bibinfo{journal}{Advances in Physics}}
  \textbf{\bibinfo{volume}{69}}, \bibinfo{pages}{249--435}
  (\bibinfo{year}{2020}).

\bibitem{BergholtzRMP2021}
\bibinfo{author}{Bergholtz, E.~J.}, \bibinfo{author}{Budich, J.~C.} \&
  \bibinfo{author}{Kunst, F.~K.}
\newblock \bibinfo{title}{Exceptional topology of non-hermitian systems}.
\newblock \emph{\bibinfo{journal}{Rev. Mod. Phys.}}
  \textbf{\bibinfo{volume}{93}}, \bibinfo{pages}{015005}
  (\bibinfo{year}{2021}).

\bibitem{Fradkin2013field}
\bibinfo{author}{Fradkin, E.}
\newblock \emph{\bibinfo{title}{Field theories of condensed matter physics}}
  (\bibinfo{publisher}{Cambridge University Press}, \bibinfo{year}{2013}).

\bibitem{Rachel2018interacting}
\bibinfo{author}{Rachel, S.}
\newblock \bibinfo{title}{Interacting topological insulators: a review}.
\newblock \emph{\bibinfo{journal}{Reports on Progress in Physics}}
  \textbf{\bibinfo{volume}{81}}, \bibinfo{pages}{116501}
  (\bibinfo{year}{2018}).

\bibitem{Andronov1937}
\bibinfo{author}{Andronov, A.~A.} \& \bibinfo{author}{Pontryagin, L.~S.}
\newblock \bibinfo{title}{Syst\`emes grossiers}.
\newblock \emph{\bibinfo{journal}{Doklady Akademii Nauk SSSR}}
  \textbf{\bibinfo{volume}{14}}, \bibinfo{pages}{247--250}
  (\bibinfo{year}{1937}).

\bibitem{Oshemkov1998}
\bibinfo{author}{Oshemkov, A.~A.} \& \bibinfo{author}{Sharko, V.~V.}
\newblock \bibinfo{title}{Classification of morse-smale flows on
  two-dimensional manifolds}.
\newblock \emph{\bibinfo{journal}{Sb. Math.}} \textbf{\bibinfo{volume}{189}},
  \bibinfo{pages}{1205} (\bibinfo{year}{1998}).

\bibitem{morse2007topology}
\bibinfo{author}{Morse, M.}
\newblock \bibinfo{title}{Topology and equilibria}.
\newblock \emph{\bibinfo{journal}{The American Mathematical Monthly}}
  \textbf{\bibinfo{volume}{114}}, \bibinfo{pages}{819--834}
  (\bibinfo{year}{2007}).

\bibitem{palis2012geometric}
\bibinfo{author}{Palis, J.~J.} \& \bibinfo{author}{De~Melo, W.}
\newblock \emph{\bibinfo{title}{Geometric theory of dynamical systems: an
  introduction}} (\bibinfo{publisher}{Springer Science \& Business Media},
  \bibinfo{year}{2012}).

\bibitem{gunther2021introduction}
\bibinfo{author}{G{\"u}nther, T.} \& \bibinfo{author}{Baeza~Rojo, I.}
\newblock \bibinfo{title}{Introduction to vector field topology}.
\newblock In \emph{\bibinfo{booktitle}{Topological Methods in Data Analysis and
  Visualization VI: Theory, Applications, and Software}},
  \bibinfo{pages}{289--326} (\bibinfo{organization}{Springer},
  \bibinfo{year}{2021}).

\bibitem{Longcope_topological_2005}
\bibinfo{author}{Longcope, D.~W.}
\newblock \bibinfo{title}{Topological {Methods} for the {Analysis} of {Solar}
  {Magnetic} {Fields}}.
\newblock \emph{\bibinfo{journal}{Living Reviews in Solar Physics}}
  \textbf{\bibinfo{volume}{2}} (\bibinfo{year}{2005}).

\bibitem{Carlsson2020}
\bibinfo{author}{Carlsson, G.}
\newblock \bibinfo{title}{Topological methods for data modelling}.
\newblock \emph{\bibinfo{journal}{Nat. Rev. Phys.}}
  \textbf{\bibinfo{volume}{2}}, \bibinfo{pages}{697--708}
  (\bibinfo{year}{2020}).

\bibitem{hasan2010colloquium}
\bibinfo{author}{Hasan, M.~Z.} \& \bibinfo{author}{Kane, C.~L.}
\newblock \bibinfo{title}{Colloquium: topological insulators}.
\newblock \emph{\bibinfo{journal}{Rev. Mod. Phys.}}
  \textbf{\bibinfo{volume}{82}}, \bibinfo{pages}{3045} (\bibinfo{year}{2010}).

\bibitem{bradlyn2017topological}
\bibinfo{author}{Bradlyn, B.} \emph{et~al.}
\newblock \bibinfo{title}{Topological quantum chemistry}.
\newblock \emph{\bibinfo{journal}{Nature}} \textbf{\bibinfo{volume}{547}},
  \bibinfo{pages}{298--305} (\bibinfo{year}{2017}).

\bibitem{Shah2024RMP}
\bibinfo{author}{Shah, T.}, \bibinfo{author}{Brendel, C.},
  \bibinfo{author}{Peano, V.} \& \bibinfo{author}{Marquardt, F.}
\newblock \bibinfo{title}{Colloquium: Topologically protected transport in
  engineered mechanical systems}.
\newblock \emph{\bibinfo{journal}{Rev. Mod. Phys.}}
  \textbf{\bibinfo{volume}{96}}, \bibinfo{pages}{021002}
  (\bibinfo{year}{2024}).

\bibitem{Cooper2019RMP}
\bibinfo{author}{Cooper, N.~R.}, \bibinfo{author}{Dalibard, J.} \&
  \bibinfo{author}{Spielman, I.~B.}
\newblock \bibinfo{title}{Topological bands for ultracold atoms}.
\newblock \emph{\bibinfo{journal}{Rev. Mod. Phys.}}
  \textbf{\bibinfo{volume}{91}}, \bibinfo{pages}{015005}
  (\bibinfo{year}{2019}).

\bibitem{Slim2024BKC}
\bibinfo{author}{Slim, J.~J.} \emph{et~al.}
\newblock \bibinfo{title}{Optomechanical realization of the bosonic kitaev
  chain}.
\newblock \emph{\bibinfo{journal}{Nature}} \textbf{\bibinfo{volume}{627}},
  \bibinfo{pages}{767--771} (\bibinfo{year}{2024}).

\bibitem{Altland2010book}
\bibinfo{author}{Altland, A.} \& \bibinfo{author}{Simons, B.~D.}
\newblock \emph{\bibinfo{title}{Condensed matter field theory}}
  (\bibinfo{publisher}{Cambridge university press}, \bibinfo{year}{2010}).

\bibitem{lado2019topological}
\bibinfo{author}{Lado, J.~L.} \& \bibinfo{author}{Zilberberg, O.}
\newblock \bibinfo{title}{Topological spin excitations in harper-heisenberg
  spin chains}.
\newblock \emph{\bibinfo{journal}{Phys. Rev. Res.}}
  \textbf{\bibinfo{volume}{1}}, \bibinfo{pages}{033009} (\bibinfo{year}{2019}).

\bibitem{Xia2020}
\bibinfo{author}{Xia, S.} \emph{et~al.}
\newblock \bibinfo{title}{Nontrivial coupling of light into a defect: the
  interplay of nonlinearity and topology}.
\newblock \emph{\bibinfo{journal}{Light: Science \& Applications}}
  \textbf{\bibinfo{volume}{9}}, \bibinfo{pages}{147} (\bibinfo{year}{2020}).

\bibitem{Mukherjee2020}
\bibinfo{author}{Mukherjee, S.} \& \bibinfo{author}{Rechtsman, M.~C.}
\newblock \bibinfo{title}{Observation of floquet solitons in a topological
  bandgap}.
\newblock \emph{\bibinfo{journal}{Science}} \textbf{\bibinfo{volume}{368}},
  \bibinfo{pages}{856--859} (\bibinfo{year}{2020}).

\bibitem{Mittal2021}
\bibinfo{author}{Mittal, S.}, \bibinfo{author}{Moille, G.},
  \bibinfo{author}{Srinivasan, K.}, \bibinfo{author}{Chembo, Y.~K.} \&
  \bibinfo{author}{Hafezi, M.}
\newblock \bibinfo{title}{Topological frequency combs and nested temporal
  solitons}.
\newblock \emph{\bibinfo{journal}{Nature Physics}}
  \textbf{\bibinfo{volume}{17}}, \bibinfo{pages}{1169--1176}
  (\bibinfo{year}{2021}).

\bibitem{Mostaan2022}
\bibinfo{author}{Mostaan, N.}, \bibinfo{author}{Grusdt, F.} \&
  \bibinfo{author}{Goldman, N.}
\newblock \bibinfo{title}{Quantized topological pumping of solitons in
  nonlinear photonics and ultracold atomic mixtures}.
\newblock \emph{\bibinfo{journal}{Nature Communications}}
  \textbf{\bibinfo{volume}{13}}, \bibinfo{pages}{5997} (\bibinfo{year}{2022}).

\bibitem{Agarwal2008}
\bibinfo{author}{Agarwal, M.} \emph{et~al.}
\newblock \bibinfo{title}{{A study of electrostatic force nonlinearities in
  resonant microstructures}}.
\newblock \emph{\bibinfo{journal}{Applied Physics Letters}}
  \textbf{\bibinfo{volume}{92}}, \bibinfo{pages}{104106}
  (\bibinfo{year}{2008}).

\bibitem{dumont2024hamiltonian}
\bibinfo{author}{Dumont, V.} \emph{et~al.}
\newblock \bibinfo{title}{Hamiltonian reconstruction via ringdown dynamics}.
\newblock \emph{\bibinfo{journal}{arXiv:2403.00102}}  (\bibinfo{year}{2024}).

\bibitem{shen1984principles}
\bibinfo{author}{Shen, Y.-R.}
\newblock \emph{\bibinfo{title}{Principles of nonlinear optics}}
  (\bibinfo{publisher}{Wiley-Interscience, New York, NY, USA},
  \bibinfo{year}{1984}).

\bibitem{dykman2012fluctuating}
\bibinfo{author}{Dykman, M.}
\newblock \emph{\bibinfo{title}{Fluctuating nonlinear oscillators: from
  nanomechanics to quantum superconducting circuits}}
  (\bibinfo{publisher}{Oxford University Press}, \bibinfo{year}{2012}).

\bibitem{Leuch2016}
\bibinfo{author}{Leuch, A.} \emph{et~al.}
\newblock \bibinfo{title}{Parametric symmetry breaking in a nonlinear
  resonator}.
\newblock \emph{\bibinfo{journal}{Phys. Rev. Lett.}}
  \textbf{\bibinfo{volume}{117}}, \bibinfo{pages}{214101}
  (\bibinfo{year}{2016}).

\bibitem{heugel2019classical}
\bibinfo{author}{Heugel, T.~L.}, \bibinfo{author}{Oscity, M.},
  \bibinfo{author}{Eichler, A.}, \bibinfo{author}{Zilberberg, O.} \&
  \bibinfo{author}{Chitra, R.}
\newblock \bibinfo{title}{Classical many-body time crystals}.
\newblock \emph{\bibinfo{journal}{Phys. Rev. Lett.}}
  \textbf{\bibinfo{volume}{123}}, \bibinfo{pages}{124301}
  (\bibinfo{year}{2019}).

\bibitem{parametricBook}
\bibinfo{author}{Eichler, A.} \& \bibinfo{author}{Zilberberg, O.}
\newblock \emph{\bibinfo{title}{Classical and quantum parametric phenomena.}}
  (\bibinfo{publisher}{Oxford University Press}, \bibinfo{year}{2023}).

\bibitem{chitra2015dynamical}
\bibinfo{author}{Chitra, R.} \& \bibinfo{author}{Zilberberg, O.}
\newblock \bibinfo{title}{Dynamical many-body phases of the parametrically
  driven, dissipative dicke model}.
\newblock \emph{\bibinfo{journal}{Phys. Rev. A}} \textbf{\bibinfo{volume}{92}},
  \bibinfo{pages}{023815} (\bibinfo{year}{2015}).

\bibitem{Soriente2021}
\bibinfo{author}{Soriente, M.}, \bibinfo{author}{Heugel, T.~L.},
  \bibinfo{author}{Omiya, K.}, \bibinfo{author}{Chitra, R.} \&
  \bibinfo{author}{Zilberberg, O.}
\newblock \bibinfo{title}{Distinctive class of dissipation-induced phase
  transitions and their universal characteristics}.
\newblock \emph{\bibinfo{journal}{Phys. Rev. Res.}}
  \textbf{\bibinfo{volume}{3}}, \bibinfo{pages}{023100} (\bibinfo{year}{2021}).

\bibitem{ferri2021emerging}
\bibinfo{author}{Ferri, F.} \emph{et~al.}
\newblock \bibinfo{title}{Emerging dissipative phases in a superradiant quantum
  gas with tunable decay}.
\newblock \emph{\bibinfo{journal}{Phys. Rev. X}} \textbf{\bibinfo{volume}{11}},
  \bibinfo{pages}{041046} (\bibinfo{year}{2021}).

\bibitem{mivehvar2021cavity}
\bibinfo{author}{Mivehvar, F.}, \bibinfo{author}{Piazza, F.},
  \bibinfo{author}{Donner, T.} \& \bibinfo{author}{Ritsch, H.}
\newblock \bibinfo{title}{Cavity qed with quantum gases: new paradigms in
  many-body physics}.
\newblock \emph{\bibinfo{journal}{Advances in Physics}}
  \textbf{\bibinfo{volume}{70}}, \bibinfo{pages}{1--153}
  (\bibinfo{year}{2021}).

\bibitem{hartmann_quantum_2008}
\bibinfo{author}{Hartmann, M.~J.}, \bibinfo{author}{Brandao, F. G. S.~L.} \&
  \bibinfo{author}{Plenio, M.~B.}
\newblock \bibinfo{title}{Quantum {Many}-{Body} {Phenomena} in {Coupled}
  {Cavity} {Arrays}}.
\newblock \emph{\bibinfo{journal}{Laser \& Photonics Reviews}}
  \textbf{\bibinfo{volume}{2}}, \bibinfo{pages}{527--556}
  (\bibinfo{year}{2008}).

\bibitem{Ritsch2013RMP}
\bibinfo{author}{Ritsch, H.}, \bibinfo{author}{Domokos, P.},
  \bibinfo{author}{Brennecke, F.} \& \bibinfo{author}{Esslinger, T.}
\newblock \bibinfo{title}{Cold atoms in cavity-generated dynamical optical
  potentials}.
\newblock \emph{\bibinfo{journal}{Rev. Mod. Phys.}}
  \textbf{\bibinfo{volume}{85}}, \bibinfo{pages}{553--601}
  (\bibinfo{year}{2013}).

\bibitem{wolf2020}
\bibinfo{author}{Wolf, G.~W.}
\newblock \bibinfo{title}{Surfaces—topography and topology}.
\newblock \emph{\bibinfo{journal}{Surf. Topogr.: Metrol. Prop.}}
  \textbf{\bibinfo{volume}{8}}, \bibinfo{pages}{014003} (\bibinfo{year}{2020}).

\bibitem{Shaw2016MEMS}
\bibinfo{author}{Polunin, P.~M.}, \bibinfo{author}{Yang, Y.},
  \bibinfo{author}{Dykman, M.~I.}, \bibinfo{author}{Kenny, T.~W.} \&
  \bibinfo{author}{Shaw, S.~W.}
\newblock \bibinfo{title}{Characterization of mems resonator nonlinearities
  using the ringdown response}.
\newblock \emph{\bibinfo{journal}{Journal of Microelectromechanical Systems}}
  \textbf{\bibinfo{volume}{25}}, \bibinfo{pages}{297--303}
  (\bibinfo{year}{2016}).

\bibitem{Bartolo2016}
\bibinfo{author}{Bartolo, N.}, \bibinfo{author}{Minganti, F.},
  \bibinfo{author}{Casteels, W.} \& \bibinfo{author}{Ciuti, C.}
\newblock \bibinfo{title}{Exact steady state of a kerr resonator with one- and
  two-photon driving and dissipation: Controllable wigner-function
  multimodality and dissipative phase transitions}.
\newblock \emph{\bibinfo{journal}{Phys. Rev. A}} \textbf{\bibinfo{volume}{94}},
  \bibinfo{pages}{033841} (\bibinfo{year}{2016}).

\bibitem{heugel2019quantum}
\bibinfo{author}{Heugel, T.~L.}, \bibinfo{author}{Biondi, M.},
  \bibinfo{author}{Zilberberg, O.} \& \bibinfo{author}{Chitra, R.}
\newblock \bibinfo{title}{Quantum transducer using a parametric
  driven-dissipative phase transition}.
\newblock \emph{\bibinfo{journal}{Phys. Rev. Lett.}}
  \textbf{\bibinfo{volume}{123}}, \bibinfo{pages}{173601}
  (\bibinfo{year}{2019}).

\bibitem{beaulieu2023observation}
\bibinfo{author}{Beaulieu, G.} \emph{et~al.}
\newblock \bibinfo{title}{Observation of first-and second-order dissipative
  phase transitions in a two-photon driven kerr resonator}.
\newblock \emph{\bibinfo{journal}{arXiv preprint arXiv:2310.13636}}
  (\bibinfo{year}{2023}).

\bibitem{kovsata2022fixing}
\bibinfo{author}{Ko{\v{s}}ata, J.}, \bibinfo{author}{Leuch, A.},
  \bibinfo{author}{K{\"a}stli, T.} \& \bibinfo{author}{Zilberberg, O.}
\newblock \bibinfo{title}{Fixing the rotating-wave approximation for strongly
  detuned quantum oscillators}.
\newblock \emph{\bibinfo{journal}{Phys. Rev. Res.}}
  \textbf{\bibinfo{volume}{4}}, \bibinfo{pages}{033177} (\bibinfo{year}{2022}).

\bibitem{seibold2024floquet}
\bibinfo{author}{Seibold, K.}, \bibinfo{author}{Ameye, O.} \&
  \bibinfo{author}{Zilberberg, O.}
\newblock \bibinfo{title}{Floquet expansion by counting pump photons}.
\newblock \emph{\bibinfo{journal}{arXiv preprint arXiv:2404.09704}}
  (\bibinfo{year}{2024}).

\bibitem{Soriente2020}
\bibinfo{author}{Soriente, M.}, \bibinfo{author}{Chitra, R.} \&
  \bibinfo{author}{Zilberberg, O.}
\newblock \bibinfo{title}{Distinguishing phases using the dynamical response of
  driven-dissipative light-matter systems}.
\newblock \emph{\bibinfo{journal}{Phys. Rev. A}}
  \textbf{\bibinfo{volume}{101}}, \bibinfo{pages}{023823}
  (\bibinfo{year}{2020}).

\bibitem{huber2020spectral}
\bibinfo{author}{Huber, J.~S.} \emph{et~al.}
\newblock \bibinfo{title}{Spectral evidence of squeezing of a weakly damped
  driven nanomechanical mode}.
\newblock \emph{\bibinfo{journal}{Physical Review X}}
  \textbf{\bibinfo{volume}{10}}, \bibinfo{pages}{021066}
  (\bibinfo{year}{2020}).

\bibitem{senthil2015symmetry}
\bibinfo{author}{Senthil, T.}
\newblock \bibinfo{title}{Symmetry-protected topological phases of quantum
  matter}.
\newblock \emph{\bibinfo{journal}{Annu. Rev. Condens. Matter Phys.}}
  \textbf{\bibinfo{volume}{6}}, \bibinfo{pages}{299--324}
  (\bibinfo{year}{2015}).

\bibitem{boness2024resonant}
\bibinfo{author}{Bone{\ss}, D.}, \bibinfo{author}{Belzig, W.} \&
  \bibinfo{author}{Dykman, M.}
\newblock \bibinfo{title}{Resonant-force induced symmetry breaking in a quantum
  parametric oscillator}.
\newblock \emph{\bibinfo{journal}{arXiv preprint arXiv:2405.02706}}
  (\bibinfo{year}{2024}).

\bibitem{Papariello2016}
\bibinfo{author}{Papariello, L.}, \bibinfo{author}{Zilberberg, O.},
  \bibinfo{author}{Eichler, A.} \& \bibinfo{author}{Chitra, R.}
\newblock \bibinfo{title}{Ultrasensitive hysteretic force sensing with
  parametric nonlinear oscillators}.
\newblock \emph{\bibinfo{journal}{Phys. Rev. E}} \textbf{\bibinfo{volume}{94}},
  \bibinfo{pages}{022201} (\bibinfo{year}{2016}).

\bibitem{Flynn2020}
\bibinfo{author}{Flynn, V.~P.}, \bibinfo{author}{Cobanera, E.} \&
  \bibinfo{author}{Viola, L.}
\newblock \bibinfo{title}{Deconstructing effective non-hermitian dynamics in
  quadratic bosonic hamiltonians}.
\newblock \emph{\bibinfo{journal}{New Journal of Physics}}
  \textbf{\bibinfo{volume}{22}}, \bibinfo{pages}{083004}
  (\bibinfo{year}{2020}).

\bibitem{delPino2022NH}
\bibinfo{author}{del Pino, J.}, \bibinfo{author}{Slim, J.~J.} \&
  \bibinfo{author}{Verhagen, E.}
\newblock \bibinfo{title}{Non-hermitian chiral phononics through
  optomechanically induced squeezing}.
\newblock \emph{\bibinfo{journal}{Nature}} \textbf{\bibinfo{volume}{606}},
  \bibinfo{pages}{82--87} (\bibinfo{year}{2022}).

\bibitem{Dai2024}
\bibinfo{author}{Dai, T.} \emph{et~al.}
\newblock \bibinfo{title}{Non-hermitian topological phase transitions
  controlled by nonlinearity}.
\newblock \emph{\bibinfo{journal}{Nature Physics}}
  \textbf{\bibinfo{volume}{20}}, \bibinfo{pages}{101--108}
  (\bibinfo{year}{2024}).

\bibitem{Heugel2024}
\bibinfo{author}{Heugel, T.~L.}, \bibinfo{author}{Chitra, R.},
  \bibinfo{author}{Eichler, A.} \& \bibinfo{author}{Zilberberg, O.}
\newblock \bibinfo{title}{Proliferation of unstable states and their impact on
  stochastic out-of-equilibrium dynamics in two coupled kerr parametric
  oscillators}.
\newblock \emph{\bibinfo{journal}{Phys. Rev. E}}
  \textbf{\bibinfo{volume}{109}}, \bibinfo{pages}{064308}
  (\bibinfo{year}{2024}).

\bibitem{Markoviv2024}
\bibinfo{author}{Markovi{\'c}, D.}, \bibinfo{author}{Mizrahi, A.},
  \bibinfo{author}{Querlioz, D.} \& \bibinfo{author}{Grollier, J.}
\newblock \bibinfo{title}{Physics for neuromorphic computing}.
\newblock \emph{\bibinfo{journal}{Nature Rev. Phys.}}
  \textbf{\bibinfo{volume}{2}}, \bibinfo{pages}{499--510}
  (\bibinfo{year}{2020}).

\bibitem{mcdonnell2009stochastic}
\bibinfo{author}{McDonnell, M.~D.} \& \bibinfo{author}{Abbott, D.}
\newblock \bibinfo{title}{What is stochastic resonance? definitions,
  misconceptions, debates, and its relevance to biology}.
\newblock \emph{\bibinfo{journal}{PLoS computational biology}}
  \textbf{\bibinfo{volume}{5}}, \bibinfo{pages}{e1000348}
  (\bibinfo{year}{2009}).

\bibitem{ragone2018computation}
\bibinfo{author}{Ragone, F.}, \bibinfo{author}{Wouters, J.} \&
  \bibinfo{author}{Bouchet, F.}
\newblock \bibinfo{title}{Computation of extreme heat waves in climate models
  using a large deviation algorithm}.
\newblock \emph{\bibinfo{journal}{Proceedings of the National Academy of
  Sciences}} \textbf{\bibinfo{volume}{115}}, \bibinfo{pages}{24--29}
  (\bibinfo{year}{2018}).

\bibitem{qiao2019applications}
\bibinfo{author}{Qiao, Z.}, \bibinfo{author}{Lei, Y.} \& \bibinfo{author}{Li,
  N.}
\newblock \bibinfo{title}{Applications of stochastic resonance to machinery
  fault detection: A review and tutorial}.
\newblock \emph{\bibinfo{journal}{Mechanical Systems and Signal Processing}}
  \textbf{\bibinfo{volume}{122}}, \bibinfo{pages}{502--536}
  (\bibinfo{year}{2019}).

\bibitem{Gammaitoni2002}
\bibinfo{author}{Gammaitoni, L.} \& \bibinfo{author}{Bulsara, A.~R.}
\newblock \bibinfo{title}{Noise activated nonlinear dynamic sensors}.
\newblock \emph{\bibinfo{journal}{Phys. Rev. Lett.}}
  \textbf{\bibinfo{volume}{88}}, \bibinfo{pages}{230601}
  (\bibinfo{year}{2002}).

\bibitem{rodriguez2020enhancing}
\bibinfo{author}{Rodriguez, S.~R.}
\newblock \bibinfo{title}{Enhancing the speed and sensitivity of a nonlinear
  optical sensor with noise}.
\newblock \emph{\bibinfo{journal}{Phys. Rev. App.}}
  \textbf{\bibinfo{volume}{13}}, \bibinfo{pages}{024032}
  (\bibinfo{year}{2020}).

\bibitem{Marchetti2013RMP}
\bibinfo{author}{Marchetti, M.~C.} \emph{et~al.}
\newblock \bibinfo{title}{Hydrodynamics of soft active matter}.
\newblock \emph{\bibinfo{journal}{Rev. Mod. Phys.}}
  \textbf{\bibinfo{volume}{85}}, \bibinfo{pages}{1143--1189}
  (\bibinfo{year}{2013}).

\bibitem{terhal2020towards}
\bibinfo{author}{Terhal, B.~M.}, \bibinfo{author}{Conrad, J.} \&
  \bibinfo{author}{Vuillot, C.}
\newblock \bibinfo{title}{Towards scalable bosonic quantum error correction}.
\newblock \emph{\bibinfo{journal}{Quantum Science and Technology}}
  \textbf{\bibinfo{volume}{5}}, \bibinfo{pages}{043001} (\bibinfo{year}{2020}).

\end{thebibliography}

\end{document}